\documentclass[twocolumn]{aastex62}

\newcommand{\rxte}{{\it RXTE}}

\received{}
\revised{}
\accepted{}
\submitjournal{ApJ}

\shorttitle{The radio/X-ray correlation in XRBs}
\shortauthors{Koljonen \& Russell}

\begin{document}

\title{The radio/X-ray correlation in X-ray binaries -- Insights from a hard X-ray perspective}

\correspondingauthor{K.~I.~I.~Koljonen}
\email{karri.koljonen@utu.fi}

\author[0000-0002-9677-1533]{Karri~I.~I.~Koljonen}
\affil{Finnish Centre for Astronomy with ESO (FINCA), University of Turku, V\"ais\"al\"antie 20, 21500 Piikki\"o, Finland}
\affil{Aalto University Mets\"ahovi Radio Observatory, PO Box 13000, FI-00076 Aalto, Finland}

\author{David~M.~Russell}
\affiliation{New York University Abu Dhabi, PO Box 129188, Abu Dhabi, UAE}



\begin{abstract}

The radio/X-ray correlation is one of the most important pieces of observational evidence of the disk-jet connection in accreting compact objects. However, a growing number of X-ray binaries seem to present deviations from the universal radio/X-ray correlation and the origin of these outliers are still very much debated. In previous studies, the X-ray bolometric luminosity used in the radio/X-ray correlation has been estimated using a narrow, soft X-ray band. We study how estimating the X-ray bolometric luminosity using broadband observations of X-ray binaries affects the radio/X-ray correlation. We found that the ratio between the broadband (3--200 keV) and narrowband (3--9 keV) luminosities varies between 5--10 in the hard X-ray state. Overall, the resulting radio/X-ray correlation slopes and normalizations did not present a very significant change suggesting that they are not affected greatly by observational biases but are caused by real physical effects. We found that all sources that reach high enough luminosity change their correlation slopes from the universal slope to a much steeper one. In addition, sources in the steeper radio/X-ray track show a distinct cutoff in the high energy X-ray spectrum at tens of keV. These results suggest that the accretion flow presents a morphological change at a certain critical luminosity during the outburst rise from radiatively inefficient to radiatively efficient flow that is in turn more efficient in cooling the hot accretion flow producing the hard X-ray emission. This change could also affect to the jet launching properties in these systems.

\end{abstract}



\section{Introduction} \label{introduction}

The radio/X-ray correlation \citep[e.g.][]{hannikainen98,corbel03,gallo03} is one of the most important pieces of observational evidence about the disk-jet connection in X-ray binaries (XRB). This connection implies that an increase in the mass accretion rate onto the compact object during an outburst event (resulting in an increase in the X-ray emission) results in an increase in mass loading to the jet (subsequently resulting in an increase in the radio emission). It has been shown, that the same correlation is present also in active galactic nuclei (AGN) given that the jet properties scale with the black hole mass \citep[e.g.][]{merloni03,falcke04}. 

In the early studies (see above), the radio/X-ray correlation seemed to be concentrated on a single track with a constant coefficient between the logarithmically scaled luminosities: $L_{\mathrm{radio}} \propto L_{\mathrm{X}}^{\sim0.7}$. This correlation coefficient could be explained by assuming that the X-ray emission arises from Compton scattering in advection-dominated accretion flow (ADAF; \citealt{heinz03}), or from optically thin synchrotron emission in the jet \citep{markoff03}, and the radio emission from the optically thick synchrotron emission in the jet, all moderated by the accretion rate (and mass of the compact object which, however, does not present a large effect on the correlation for stellar-mass-sized black holes). However, when more simultaneous X-ray and radio observations were taken on different sources, it became clear that serious deviations exist with some sources presenting a radio/X-ray correlation with a different coefficient and/or lower monochromatic radio luminosity at a given X-ray luminosity (or higher X-ray luminosities at a given radio luminosity). In a population study by \citet{gallo12}, two groups of XRBs were found in the radio/X-ray plane via cluster analysis: one with a coefficient of 0.6 (radio-loud) and the other with 1.0 (radio-quiet). The origin of this discrepancy is still very much debated. \citet{gallo14,gallo18} concluded that robust partitioning of the black hole XRBs to two or more groups could not be achieved collectively. However, a large intrinsic scatter of the relation allows individual sources to follow different tracks along the mean relation: e.g. H1743$-$322 shows different radio/X-ray correlation coefficient ranging from 0.0 to 1.4 depending on the luminosity \citep{coriat11}. They argued the steeper coefficient found in high luminosities to be an effect of the radiative efficiency of the accretion flow converting the accretion energy to radiation. In this case, the steeper coefficient could be achieved by assuming that the X-ray luminosity is directly proportional to the mass accretion rate, instead of being proportional to the square of the mass accretion rate as in ADAF models. On the other hand, a lower radio luminosity would be expected from sources that have stronger magnetic fields due to radiative losses \citep{casella09}, and a higher X-ray luminosity would be expected from sources where a cool, inner disk would form by re-condensation from the ADAF producing more seed photons for Comptonization \citep{meyerhofmeister14}. In addition, sources with different jet inclination angles to the line-of-sight can have an effect to the received radio emission by an order of magnitude \citep{zdziarski16,motta18}, and sources with different disk inclinations present differences in their X-ray properties \citep{munosdarias13,heil15,motta15,motta18}.           

\citet{espinasse18} studied the radio spectral properties of the two groups and found that the radio-quiet sources present steeper radio spectra (i.e. lower spectral index $\alpha_{R}$, where $F_{\nu} \propto \nu^{\alpha_{R}}$). Thus, at a usual observing frequency of a few GHz, the radio luminosity of the radio-quiet sources is lower than the radio-loud group, possibly explaining part of the discrepancy. In addition, radio-quiet sources present lower rms variability in the X-ray lightcurves than radio-loud sources \citep{dincer14}, and the two groups might differ in source inclination \citep{motta18}. In this paper, we look at the sources in the X-ray regime. In previous studies, the X-ray band used to measure the source luminosity was restricted typically to 3--9 keV, partly due to detector efficiencies and partly due to reproducibility with earlier studies. The differences in the X-ray spectra when taking into account a wider X-ray band can have an effect on the bolometric X-ray luminosity and may differ in the two groups of sources. An encouraging example was recently set by \citet{bernardini16}, who found that the bolometric X-ray luminosity correction to the optical/X-ray correlation in XRBs can explain some of the discrepancies between the black hole and neutron star systems in the optical/X-ray correlation plane. 

\section{Observations} \label{observations}

We investigated the effect of including the hard X-ray spectrum in the determination of the bolometric X-ray luminosity by selecting sources with the most simultaneous radio and X-ray (Rossi X-ray Timing Exporer; \rxte) data. This resulted in sources GX 339--4 and XTE J1118$+$480 from the radio-loud group\footnote{Whether XTE J1118$+$480 is a radio-loud source is a matter of debate, since it has never reached the X-ray luminosities where the radio-quiet/radio-loud dichotomy seems to appear.}, H1743--322 and Swift J1753.5--0127 from the radio-quiet group, and GRO J1655--40 and XTE J1752--223 for transitional sources between these groups. These sources present enough data in the archives with a scatter in both radio and X-ray luminosities that individual radio/X-ray correlation analysis is possible. 

The radio data, mostly in 4.9 GHz and/or 8.4 GHz, were obtained from the literature: \citet{mcclintock09,jonker10,coriat11,millerjones12} for H1743--322, \citet{brocksopp10} for XTE J1118$+$480, \citet{corbel13,gandhi11} for GX 339--4, \citet{soleri10} for Swift J1753.5--0127, \citet{shaposhnikov07} for GRO J1655--40, and \citet{brocksopp13} for XTE J1752--223.

We obtained (quasi-)simultaneous \rxte\/ data from the High Energy Astrophysics Science Archive Research Center (HEASARC). We reduced each pointing using \textsc{heasoft 6.22} and standard methods described in the \rxte\/ cookbook. Both the Proportional Counter Array (PCA) and the High Energy X-ray Timing Experiment (HEXTE) data were reduced and spectra were obtained from PCU--2 (all layers) and HEXTE clusters A and B (when available) for all sources. 

For spectral analysis, we grouped the data to a minimum of 5.5 sigma significance per bin, and excluded bins below 3.5 keV and above 22 keV, and below 18 keV for PCA and HEXTE, respectively. In addition, 0.5\% and 1\% systematic error were added to all channels for PCA and HEXTE, respectively. 

\section{Results} \label{results}

To study the effect of including the hard X-ray band to the X-ray luminosity measurements, we fit the broadband X-ray spectra for every source with an absorbed cutoff power law model with reflection ({\sc tbabs}; \citealt{wilms00}, {\sc pexrav}; \citealt{magdziarz95}) using the Interactive Spectral Interpretation System (ISIS; \citealt{houck00}). As PCA is not sensitive to energies below 3 keV, we could not constrain the absorption column with the fits. Thus, we fixed the absorption column to the following values found from the literature that were obtained using instruments sensitive to softer X-ray band than PCA: 0.8 for GRO J1655--40 \citep{brocksopp06,diaztrigo07}, 0.6 for GX 339--4 \citep[e.g.][]{cadollebel11}, 1.8 for H1743--322 \citep{prat09}, 0.2 for Swift J1753.5--0127 \citep{tomsick15}, 0.01 for XTE J1118$+$480 \citep{mcclintock01} and 0.6 for XTE J1752--233 \citep{chun13} in units of $10^{22}$ atoms cm$^{-2}$. We included also a Gaussian line to model the iron line at 6.4 keV when necessary. After a successful fit was obtained, we calculated the unabsorbed, model flux in the energy bands of 3--9 keV and 6--10 keV (denoted as narrowband X-ray fluxes), and 3--200 keV (denoted as broadband X-ray flux)\footnote{The flux below 3 keV can also be significant, but mostly in the softer X-ray states that are not considered in this paper.}. In addition, we corrected the fluxes for the Galactic ridge emission when necessary. The fluxes were then converted into luminosities using following distance estimates: 8.5 kpc for H1743--322 \citep{steiner12} and GX 339--4 \citep{hynes04,zdziarski04,parker16}, 1.7 kpc for J1118+480 \citep{gelino06}, 3 kpc for Swift J1753.5--0127 (see \citealt{tomsick15}, and references therein), 3.5 kpc for XTE J1752--233 \citep{shaposhnikov10}, and 3.2 kpc for GRO 1655--40 \citep{hjellming95}. 

\subsection{Narrowband vs. broadband luminosity}

\begin{figure*}
  \centering
  \includegraphics[width=1.0\linewidth]{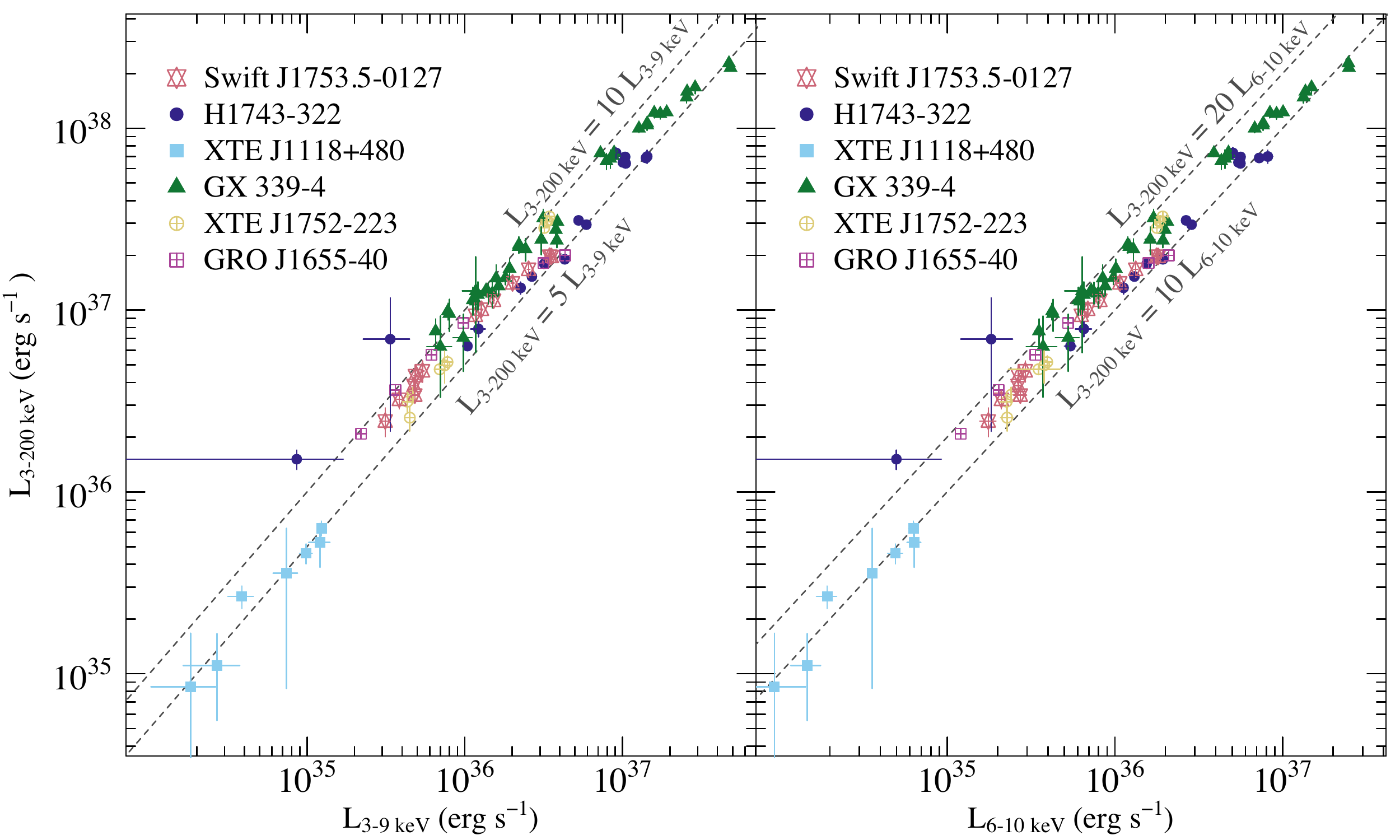}
  \caption{Comparison of the narrowband (3--9 keV and 6--10 keV) luminosities with the broadband (3--200 keV) luminosity of the sources studied in this paper. For all sources, the broadband luminosity is a factor of 5--10 or 10--20 times the 3--9 keV and 6--10 keV luminosity, respectively. However, an evolution of the factor is evident with most sources presenting a decrease of the factor with increasing luminosity. Both narrowband luminosities show similar evolution, thus we can disregard any soft component contributing to the 3--6 keV band and causing the change of the luminosity ratio.}
  \label{ff_plot}
\end{figure*}

In Fig. \ref{ff_plot}, we compare the two narrowband luminosities with the broadband luminosity showing that the broadband luminosity can be estimated to be roughly 5--10 times the 3--9 keV luminosity and 10--20 times the 6--10 keV luminosity. However, this factor varies within the source evolution in the hard state. Most sources present a higher factor for lower luminosities that gradually decreases with increasing luminosity. The exceptions are XTE J1752--223, which show an increase of the luminosity ratio with increasing luminosity, and XTE J1118$+$480, which seem to present a constant factor between the narrow- and broadband luminosities. These differences might arise from XTE J1118$+$480 being at very low luminosities compared to other sources, and XTE J1752--223 having rather poor coverage. Since both narrowband luminosities show similar evolution with the broadband luminosity, the change in the luminosity ratio is not caused by an increased soft X-ray flux in the 3--6 keV band. Rather, the difference lies in the evolution of the hard X-ray spectra.   

\begin{figure}
  \centering
  \includegraphics[width=1.0\linewidth]{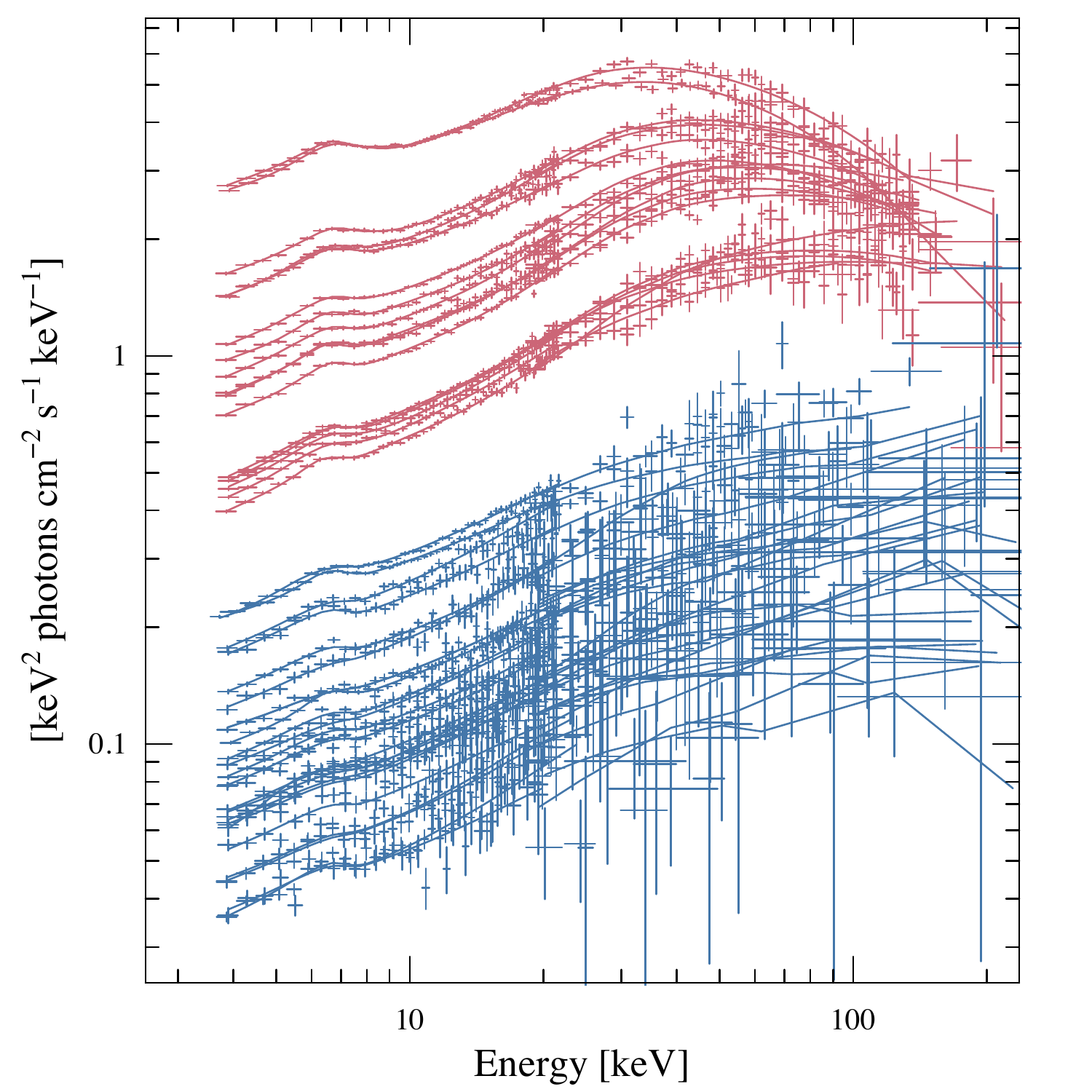}
  \caption{The joint PCA+HEXTE spectra of GX 339--4 from the same pointings that are used in Fig. \ref{ff_plot} divided into two groups: X-ray luminous group in red (with decreasing ratio between the narrow- and broadband luminosities), and less luminous group in blue (with a constant luminosity ratio). The main difference between these groups apart from the X-ray luminosity is the emergence of the hard X-ray cutoff energy ($\lesssim$200 keV) in the X-ray luminous group.}
  \label{GX339_spec}
\end{figure}

In Fig. \ref{GX339_spec}, we have plotted the joint PCA and HEXTE spectra of GX 339--4 as an example, where the data and modeling show that the peak/cutoff temperature in the hard X-ray spectrum is anticorrelated with the luminosity. The spectra colored in red correspond to the pointings where the luminosity ratio starts to decrease at a broadband luminosity exceeding $5\times10^{37}$ erg/s. Thus, the increasing hard X-ray spectral curvature is causing the decrease in the luminosity ratio. This shows that in the bright, hard X-ray state the narrowband luminosity is not always a good proxy for the bolometric luminosity as it does not take into account the hard X-ray spectral evolution. In addition, using the broadband X-ray luminosity will necessarily affect to the radio/X-ray correlations measured for the sources which we will quantify in the next section.

\subsection{Radio/broadband X-ray correlation}

\begin{figure*}
  \centering
  \includegraphics[width=1.0\linewidth]{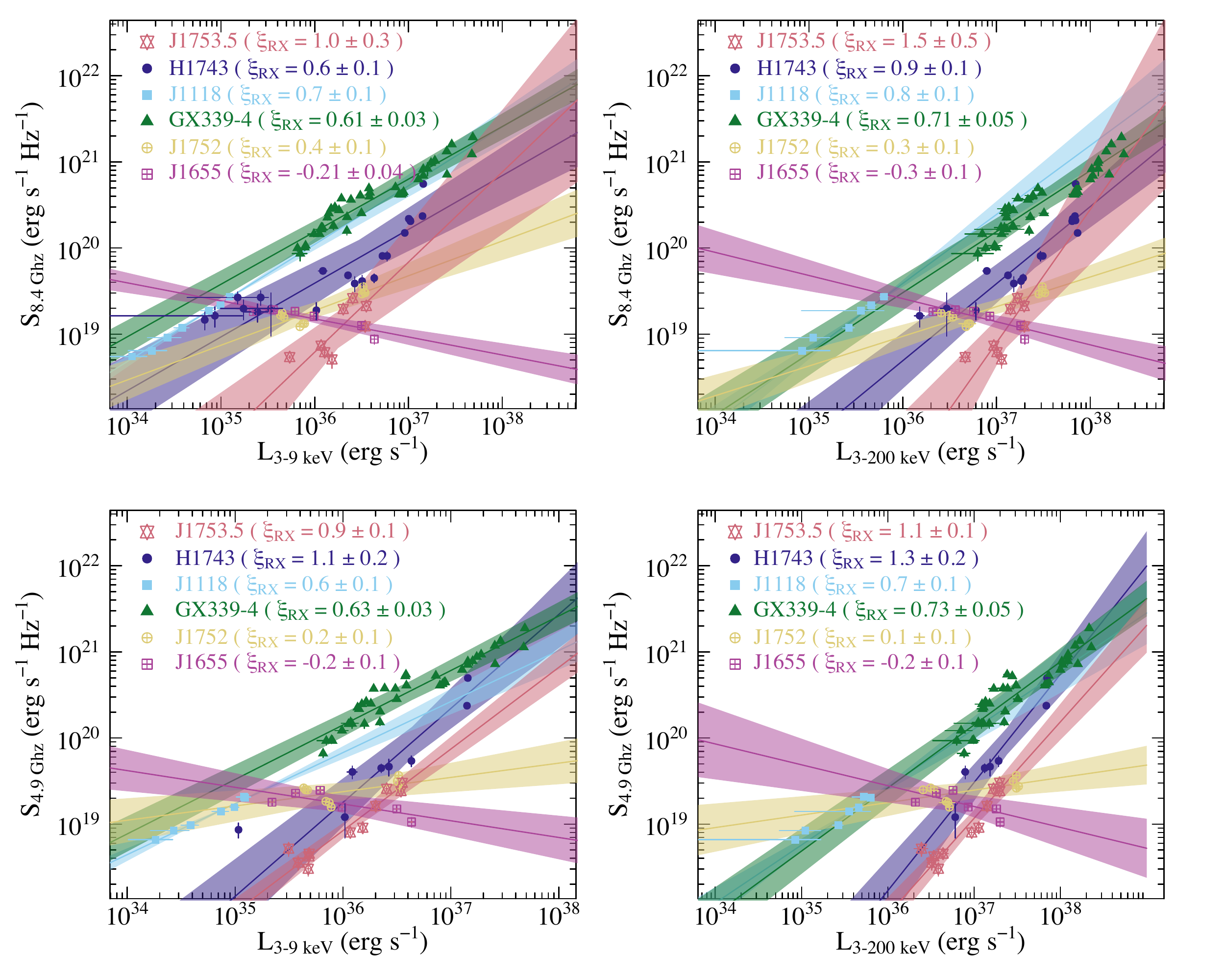}
  \caption{The radio/X-ray plane including two radio-loud sources (GX 339--4 and XTE J1118$+$480), two radio-quiet sources (H1743--322 and Swift J1753.5--0127) and two transitional sources, i.e. in between the two tracks formed by the radio-loud and radio-quiet sources (XTE J1752--233 and GRO 1655--40). The panels on the left show the radio/X-ray plane using the ``classical'' X-ray luminosity calculated from the 3--9 keV band, while the panels on the right show the plane using the X-ray luminosity calculated from the 3--200 keV band. We have used two single frequency radio flux measurements: 8.4 GHz in the upper panels and 4.9 GHz in the lower panels. The best-fit line is plotted for each source in the panels with 1$\sigma$ error contours including the intrinsic model scatter. The slopes for the best-fit lines and their 1$\sigma$ errors are marked on the panels as well.}
  \label{fp_plot}
\end{figure*}

To construct the radio/X-ray correlations, we searched for quasi-simultaneous radio observations in the literature (see Section \ref{observations}) and selected those which were observed within a day of the X-ray observations. On average, the radio observations were observed within 10 hours of the X-ray observations, with 50\% within 6.6 hours and 75\% within 16 hours. In Fig. \ref{fp_plot}, we have plotted the 3--9 keV (left column) and 3--200 keV (right column) X-ray luminosities against the quasi-simultaneous, single frequency radio luminosities in 8.4 GHz (top row) and 4.9 GHz (bottom row).

For determining the radio/X-ray correlation coefficients, we calculated a linear regression of the logarithmic radio and X-ray luminosities from individual sources. Because both variables are measured quantities, we minimized the slope and the normalization of the regression in both directions. We used the \textsc{hyper.fit} package \citep{robotham15} to fit a line to the data using the maximum likelihood method including the intrinsic scatter of the model, and to estimate the errors on the slope and normalization. This method assumes that the data errors are Gaussian, the sample is drawn randomly from the model population and the intrinsic scatter is Gaussian. We also added corrections to the intrinsic scatter due to the small sample sizes as described in the Appendix A of \citet{robotham15}. With the above assumptions, we show the best fitting model lines in Fig. \ref{fp_plot} with the shaded areas representing 1$\sigma$ errors on the slope and the intercept in addition to the intrinsic scatter. The 1$\sigma$ error values on the slope are shown also in the panel legend for individual sources.

Using the broadband instead of the narrowband X-ray luminosity resulted in similar coefficients for GX 339--4 and XTE J1118$+$480 (i.e. the radio-loud group) for both single frequency radio luminosities. For these sources, the slopes of the best fit lines are consistent with being $\xi_{RX} = 0.7-0.8$. For the radio-quiet group, the slopes are slightly steeper. The 67\% confidence limits (1$\sigma$) on the slope using broadband X-ray flux for H1743--322 are $\xi_{RX} = 0.8-1.0$ and $\xi_{RX} = 1.1-1.5$ for 8.4 and 4.9 GHz single frequency radio luminosities, respectively. The correlation slope is not well-defined for Swift J1753.5--0127 when using the 8.4 GHz radio data, but for 4.9 GHz radio data the 67\% confidence limits are $\xi_{RX} = 1.0-1.2$. For sources XTE J1752--233 and GRO 1655--40, that are found in between the radio-quiet and radio-loud groups, the slopes are very similar for both luminosity ranges but flatter for 4.9 GHz radio data.  

In a broad sense, estimating the X-ray luminosity more accurately in taking into account the hard X-ray emission did not present a major effect on the radio/X-ray correlation. In the right panels of Fig. \ref{fp_plot}, GX 339--4 and J1118$+$480 (the two radio-loud sources) form a well-defined correlation with a correlation index of 0.7--0.8, and all the radio-quiet sources lie below that correlation by about an order of magnitude. This factor of $\sim$10 difference is about the same when using the narrowband luminosity. Thus, the bolometric correction cannot explain the systematic radio-loud/radio-quiet dichotomy. 

\subsection{Radio/X-ray correlation slopes}

Overall, using the broadband X-ray luminosity presents a slight increase in the correlation coefficients (steepening of the slopes). In addition, it seems that the correlation slopes for the radio-quiet group are steeper than the radio-loud group. However, the radio/X-ray correlation slope for the X-ray luminous group (corresponds to an outburst rise) of the radio-loud source GX 339--4 is, in fact, more similar to the one for the radio-quiet group than the one calculated from the whole GX 339--4 sample when taking into account the broadband X-ray flux. Recently, \citet{islam18} came to the same conclusion. The more luminous group aligns (in both slope and normalization) with the radio-quiet sources (this can be best seen in the bottom-right panel of Fig. \ref{fp_plot}, where the correlation coefficient is $\sim$1.1 when using only the X-ray brighter group, i.e. above $5\times10^{37}$ erg/s). This can imply that all sources could transit between the two different radio/X-ray tracks, depending on the X-ray luminosity. 

\begin{figure}
  \centering
  \includegraphics[width=1.0\linewidth]{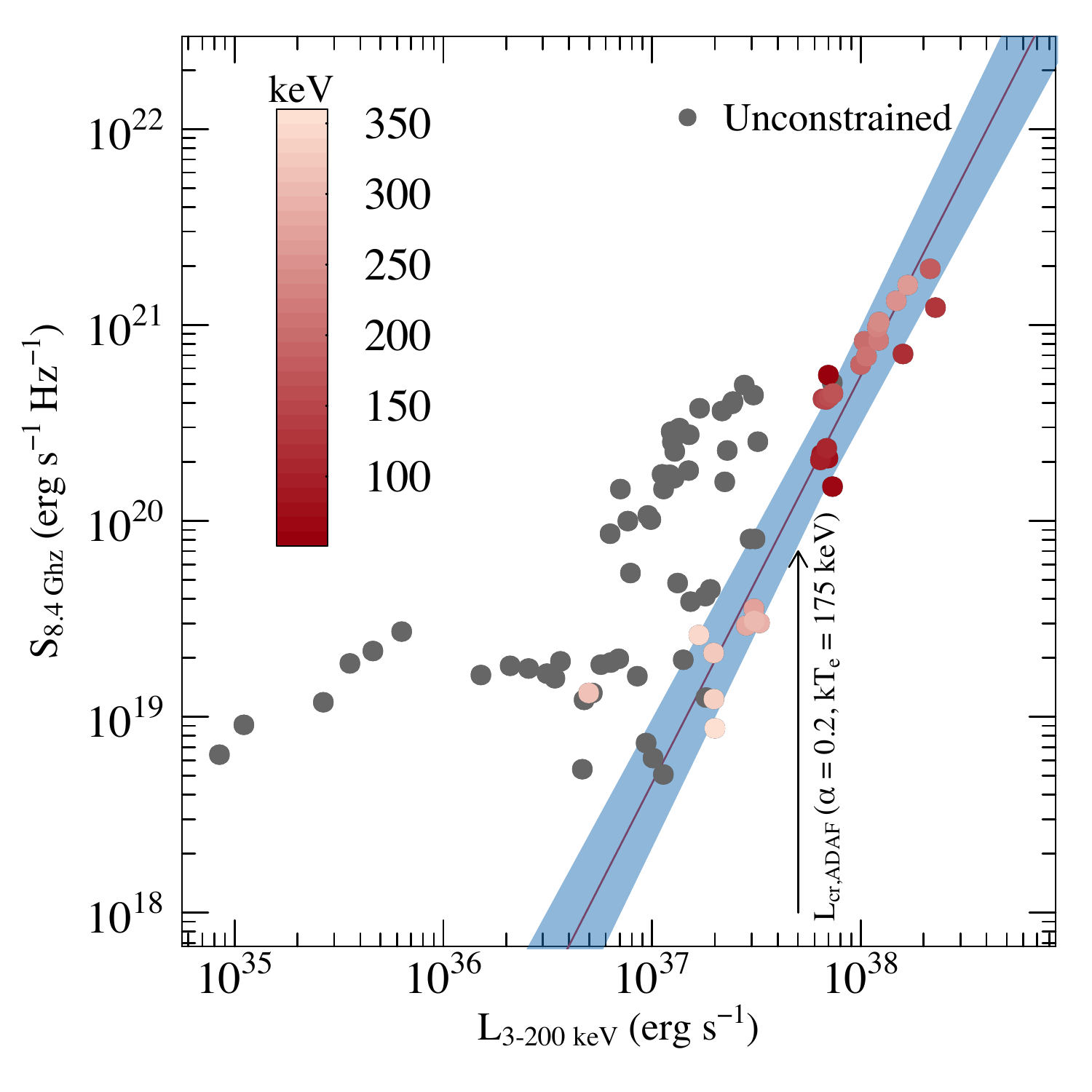}
  \caption{Radio/X-ray correlation (Fig. \ref{fp_plot}, top right) with the hard X-ray cutoff colored as a hue of red depending on the energy and grey if it is unconstrained. The sources show a measurable cutoff when they are aligned in the steeper track (red line, slope 2.1$\pm$0.1, 1$\sigma$ uncertainties plotted in blue). A critical luminosity for an efficiency change in the accretion disk for one pair of parameters suitable for GX 339--4 is shown as an arrow (see discussion in Section \ref{XX}).}
  \label{Ecut}
\end{figure}

The change to the steeper slope in GX 339--4 corresponds to the time when the source starts to shift from the luminosity ratio $L_{3-200 keV}/L_{3-9 keV} = 10$ to $L_{3-200 keV}/L_{3-9 keV} = 5$ (Fig. \ref{ff_plot}), and where the cutoff in the hard X-ray spectra appears (red group in Fig. \ref{GX339_spec}). In a similar fashion, we found that in the steeper track the hard X-ray spectra show similar morphology in all sources presenting a distinct hard X-ray cutoff below $\sim$300 keV, while elsewhere the cutoff is unconstrained meaning that the 90\% confidence range for the cutoff energy pegs to 1000 keV in the spectral fitting and lies over the passband of HEXTE (Fig. \ref{Ecut}). Assuming that the spectrum arises from thermal Comptonization, this implies that the electrons producing the Comptonized emission are effectively cooler in the steeper track, which implies a morphological change in the accretion flow. The line fitted to the data points that present a measurable cutoff in the X-ray spectrum has a coefficient of $2.1\pm0.1$ (red line in Fig. \ref{Ecut}; the coefficient is $1.9\pm0.1$ when using 4.9 GHz radio data), that is steeper than the radio/X-ray correlation slope in any single source but could indicate the maximum amount of efficiency attainable by an XRB. Interestingly, one of the highest radio/X-ray correlation coefficient found among XRBs: $L_{R} \propto L_{X}^{1.8\pm0.2}$ from MAXI J1836--194, could be then accommodated with varying accretion efficiency instead of the proposed scenario of varying jet Lorentz factor \citep{russell15}.

\section{Discussion} \label{discussion}

To summarize the above results, we have found that the bolometric correction is not constant but decreases for higher luminosities, due to the evolving hard X-ray spectrum. However, it cannot explain the radio-loud/radio-quiet dichotomy. In addition, the appearance of a cutoff in the hard X-ray spectra is linked to the source being in the steeper track in the radio/X-ray plane, that seems to also apply to an radio-loud source GX 339--4. In the following, we will discuss these findings in more detail from three perspectives: the radio-loud/radio-quiet dichotomy is mainly based on differences found in the radio properties, or on differences in the X-ray properties, or is a geometrical effect.

\subsection{Radio-loud/radio-quiet paradigm}

In a recent paper, \citet{espinasse18} studied the distribution of radio spectral indices in the radio/X-ray correlation. They found significant differences in the two populations of XRBs, such that the radio-quiet population presented steeper radio spectra as an average. The difference in the peaks of the distributions is $\sim0.4$, with radio-quiet and radio-loud sources presenting radio spectral indices $\alpha_{RQ} \sim -0.2$ and $\alpha_{RL} \sim 0.2$, respectively. In the framework of the fundamental plane, where the correlation coefficient can be estimated as $\xi_{RX} \approx (1.42-0.67\alpha_{R})q^{-1}$, where $q$ is the accretion efficiency (see \citealt{heinz03}, their equation 12b), this means a difference in $\xi_{RX}$ of 1.3/$q$ to 1.6/$q$. To relate this to the observed values of coefficients for the radio-loud group ($\xi_{RX} = 0.7-0.8$) requires $q\sim1.7$. Using this value for the radio-quiet group places the slope coefficient to $\xi_{RX} = 0.9$. Thus, having a lower spectral index in the radio, the radio-quiet sources should have higher slope coefficients, and $\xi_{RX} = 0.9$ is more or less consistent with the observed indices (collectively \citet{gallo12} found that the slope for the radio-quiet sources is 0.98$\pm$0.08). It is thus possible, that both the radio-quiet and radio-loud groups differ only in normalization in the radio/X-ray plane. On the other hand, using the observables, i.e. the average radio/X-ray correlation indices for the individual radio-loud and radio-quiet sources used in this paper at 4.9 GHz ($\xi_{RX,RL} = 0.7\pm0.1$ and $\xi_{RX,RQ} = 1.2\pm0.2$, respectively) and the corresponding radio spectral indices from \citealt{espinasse18}; $\alpha_{RL} = 0.2\pm0.2$ and $\alpha_{RQ} = -0.2\pm0.3$, we can place estimates to the value of $q$ by rearranging the fundamental plane equation for $\xi_{RX}$. Thus, the observed values can be reproduced with $q_{RL} = 1.9\pm0.3$ and $q_{RQ} = 1.3\pm0.3$, indicating that the different correlation slopes and radio spectral indices between the radio-loud and radio-quiet sources could be explained by assuming different X-ray radiative efficiencies (different values of $q$) as will be discussed in the next section. 

In addition, no apparent connection between the radio flux and inclination or black hole spin were found in the study by \citet{espinasse18}. There can be a difference in the viscosity parameter from source to source, however, there seems to be no clear difference between sources that are radio-loud or radio-quiet (comparing sources in \citealt{espinasse18} to \citealt{tetarenko18}). The radiative efficiency in the accretion disk and/or in the jet can be different for the radio-loud and radio-quiet sources, however, as discussed below the former is most likely a function of the mass accretion rate and therefore contributes to the slope of the correlation. For the latter, \citet{espinasse18} speculate that there can be distinct physical differences in the jets of the radio-loud and radio-quiet sources. The radio-loud sources could represent partially self-absorbed jets with continuous energy dissipation occurring via internal shocks (e.g. \citealt{jamil10}) and the radio-quiet sources could represent jets with more discrete ejecta that are ``lit-up'' by a shock zone (similar to what have been observed from AGN, e.g. \citealt{marscher08}) which could then explain the differences in the radiative efficiency and radio spectral indices of the jet. However, this would require that the jet morphology should change when a source transits to/from the steeper track (e.g. H1743--322, XTE J1752--223, GRO 1655--40) and we would observe a shift in the radio spectral index. While double-frequency observations are scarce during the transitional phase for XRBs, GRO 1655--40 shows very little change at the radio spectral index or radio flux density during the transition with the radio spectral index staying at $\alpha_{R}>0.2$ \citep{shaposhnikov07}. H1743--322 shows a flat or inverted spectrum when the source is in the steep track, however, no information is available for the value of the radio spectral index elsewhere \citep{jonker10}. In addition, XTE 1752--223 shows both flat (during outburst rise) and optically thin (during outburst decay) spectral indices while remaining in the steep track \citep{brocksopp13}. These examples show that the radio spectral index and the single frequency radio luminosity do not vary much in the source evolution along the fundamental plane during the transitional phase between the two tracks. Thus, the difference could be found in the amount of X-ray luminosity.

\subsection{X-ray-loud/X-ray-quiet paradigm} \label{XX}

The association of the steep track with the evolution of the hard X-ray spectrum could be explained by a morphological change in the accretion flow. In this scenario, there is an increase in the accretion efficiency leading to an increased X-ray emission. The radio-quiet sources would be radiatively more efficient than the radio-loud sources. The emergence of the hard X-ray cutoff in the spectra when sources are in the radiatively efficient track implies effective cooling of the Compton upscattering electrons by a soft seed photon population.

For the radiatively inefficient accretion flows (RIAFs), it has been shown that the radiative efficiency changes with the mass accretion rate non-linearly up to a critical luminosity where the accretion flow changes from ADAF or Type I luminous hot accretion flow (LHAF) to Type II LHAF or some cold disk/corona configuration \citep{xie12}. Close to the critical luminosity, $L_{\mathrm{cr,ADAF}} \approx 5 \theta^{3/2} \alpha^{2} \dot{M}_{\mathrm{Edd}}$, where $\theta$ is the electron temperature in keV and $\alpha$ is the viscosity parameter, the efficiency exhibits an increase from 1\% to 8\% with very little change in the mass accretion rate, after which the efficiency is approximately constant. This framework has been proposed to explain the transitional track in the radio/X-ray plane, where the sources move horizontally, i.e. increasing their X-ray luminosity, while the radio luminosity stays constant \citep{xie16}. 

After crossing the critical luminosity, the accretion flow would transfer into a two-component disk/corona configuration. The cold disk would form underneath the Comptonizing material gradually cooling it down as marked by the decreasing electron temperature (Fig. \ref{Ecut}), but remains not visible until the Comptonized layer has been cooled down exposing the underlying disk and shifting the source hardness fast to the softer part of the hardness-luminosity diagram (the ``usual'' state change). Recent work by \citet{poutanen18} showed that to reproduce similar hard power law indices as observed in the hard X-ray state spectra of XRBs, more seed photons for Comptonization are needed in the form of cold clouds in the corona or cyclo-synchrotron radiation in the hot accretion flow. It has also been suggested that the accretion disk is already at the innermost stable circular orbit at the beginning of hard-to-soft state transition \citep{koljonen15}, which indicates the formation of the cold accretion disk underneath the hot accretion flow during the hard state rise. The different state transition luminosities would be then moderated by the time the Comptonizing material has been cooled down. 

Why different sources stay on the radiatively inefficient track and others transfer to the radiatively efficient track depend on the difference in their respective critical luminosities, that are a function of the temperature of the electron population and the viscosity parameter. If the electrons in the accretion flow can be kept at a hot temperature, the critical luminosity is larger, and the source can stay in the radiatively inefficient track longer. However, when the electrons start to cool, the critical luminosity drops below the current luminosity and the source shifts to the radiatively efficient track. As an example, In Fig. \ref{Ecut} we show the critical luminosity with the viscosity parameter $\alpha$=0.2 (as estimated in \citealt{tetarenko18} for GX 339--4) and electron temperature of 175 keV (corresponding to cutoff energies $\sim$350--500 keV). We find that it coincides well with the dividing luminosity of the two groups in GX 339--4.

In addition, \citet{dincer14} showed that the fractional rms variability does not reach equal strength in radio-loud and radio-quiet sources despite similar spectral parameters (power law index, luminosity) with the former presenting overall higher rms than the latter. In a similar fashion, \citet{munosdarias11} found that the fractional rms variability decreased from 40\% to 30\% with luminosity in the rising hard state of GX 339--4. This reduction of the fractional rms in both cases could be associated with a second, less variable component diluting the observed rms variability. As discussed above, the formation of a cold disk underneath the hot flow would in addition to cooling down the electrons (and producing the spectral cutoff) reduce the observed rms variability. This is another observational evidence uniting the luminous hard state in GX 339--4 with the X-ray properties of the radio-quiet sources. 

\subsection{Geometrical effects}

In many recent studies, evidence has been accumulated about inclination affecting to the X-ray observables in both spectral and timing domains. The spectral effects include harder spectra in the hard X-ray state \citep{heil15}, triangular hardness-intensity diagrams (HIDs) and hotter accretion disks \citep{munosdarias13} for high inclination sources (e.g. H1743--322), while low inclination sources (e.g. GX 339--4) present more softer spectra, ``boxy'' HIDs and cooler disks. In the timing domain, low inclination sources have higher ``hard line'' slopes \citep{motta18}, i.e. the rate of decrease in the rms variability in the hard state discussed above, and weaker/stronger amplitudes of type-C/type-B quasi-periodic oscillations \citep{motta15}. Therefore, the apparent bolometric flux evolution in the radio/X-ray plane could be different for sources with different inclinations. In fact, for anisotropic models of X-ray emission (e.g. a slab corona) a difference in the received emission can be pronounced since the optical depth of the slab increases when viewed with higher inclinations. In addition, the reflected coronal emission from the accretion disk depends on the inclination of the source \citep{petrucci01}, and if the Comptonized emission comes deep from the gravitational field of a rapidly spinning black hole, gravitational redshift will deform the intrinsic spectrum depending on the viewing angle \citep{niedzwiecki05}. Of course, evolution and/or the variability from source-to-source of the parameters of the electron population (temperature, optical depth) Compton upscattering the seed photon spectrum will also result in similar spectral effects, thus making determining the inclination effects a challenging task.    

On the other hand, inclination is expected to affect to the radio flux received by the observer from the jet due to Doppler boosting. In \citet{soleri11}, a toy model for the jet assuming that $L_{R} \propto L_{X}\delta^2$, where $\delta$ is the Doppler boosting factor, was able to produce the spread of the radio/X-ray plane with variable inclination resulting to the radio emission from the high-inclination sources being Doppler de-boosted producing lower radio emission. There are also hints that the radio loudness correlates with the orbital inclination \citep{motta18}. However, the observed evolution of the radio-quiet sources transferring to the $L_{R} \propto L_{X}^{\sim0.7}$ track cannot be explained by Doppler boosting. Also, it is not certain whether the outer disk/orbital inclination corresponds to inner disk/jet inclination, as the spin axis of the black hole probably differs from the orbital inclination \citep{maccarone02,king18}, bringing further complications to the inclination dependence of the source properties.  

All in all, it is certain that the inclination affects to both X-ray and radio luminosity in a non-linear way due to the anisotropies most likely present in the geometries of the emitting components. The magnitude of this effect and its ability to explain both the increase in the X-ray emission (or the lack of increase in the radio emission) during the transitional phase between the two tracks and the higher radio/X-ray correlation index remains to be studied in the future.         

\section{Conclusions} \label{conclusions}

We studied the effect of including a wider (3--200 keV) X-ray band in determining the broadband X-ray luminosity for the radio/X-ray correlation using (quasi-)simultaneous \rxte\/ and 4.9 or 8.4 GHz radio data. In doing so, we characterized the hard X-ray spectrum using only data where the X-ray spectrum was well measured up to 200 keV. There seems to be little difference in the soft X-ray spectral shape of the radio-loud and radio-quiet sources in the hard state, however, an anti-correlation between the high energy cutoff energy and the X-ray luminosity is present in the spectra as have been noticed previously in many sources. The high energy cutoff typically resides at $>$500 keV, above the energy range sampled, for low X-ray luminosities, but the cutoff decreases in energy to tens of keV for high X-ray luminosities in the hard X-ray state (outburst rise). This will necessarily have an effect on the broadband X-ray luminosity and the radio/X-ray correlation slope. 

We found that the ratio between the broadband and the more traditional soft X-ray band (3--9 keV) luminosity varies between 5--10 during the hard X-ray state for individual X-ray binaries, with the factor decreasing typically with the luminosity. This makes the radio/X-ray correlation coefficients of individual sources higher (slopes are steeper) if we adopt the broadband X-ray luminosity instead of the narrowband luminosity. In addition, the relative normalizations of sources on the radio/X-ray correlation do not change much when adopting the broadband luminosity. In other words, the radio-quiet sources are still radio-quiet (or X-ray loud) despite the use of the wider band. We also found that GX 339--4, previously known as one of the classical X-ray binaries in the radio/X-ray plane presenting a canonical $L_{R} \propto L_{X}^{\sim0.7}$ track, changes to a steeper track when crossing a luminosity of 2--18\% of the Eddington luminosity (this was also independently found in a recent paper by \citealt{islam18}). This luminosity coincides with the change in the hard X-ray spectra, showing a distinct cutoff below $\sim$200 keV. In addition, we found that when the sources are located in the steeper track in the radio/X-ray plane, they present a distinct high-energy cutoff in the X-ray spectrum indicating an efficient cooling in the hot accretion flow.

We proposed that the different cutoff energies (a proxy for electron temperatures) and perhaps viscous parameters in the hot accretion flow between sources result in different critical luminosities, above which the accretion flow becomes radiatively efficient. This results in higher radio/X-ray correlation coefficients at the highest luminosities, not just for radio-quiet sources such as H1743--322, but also in the radio-loud sources such as GX 339--4. These results are in line with the assumption that the accretion flow in X-ray binaries changes from advection-dominated hot accretion flow to an underlying cold disk/hot corona-type of flow at a critical luminosity corresponding to a few percents of the Eddington luminosity. The formation of the cold accretion disk underneath the hot accretion flow will result in cooling of the flow resulting in the observed evolution of the hard X-ray cutoff and reduction in the rms variability. In addition, the different state transition luminosities that are observed from X-ray binaries would be then explained by moderating the time that the Comptonizing material has been cooled down and the cold disk underneath is exposed.  We have shown that the variable X-ray spectrum could indicate a morphological change in the accretion flow, which can also affect to the jet launching and therefore radio properties from X-ray binaries. E.g., it remains to be studied whether the differences in the accretion flow could explain why radio-quiet sources have more inverted radio spectra. In our previous work, we have shown that the X-ray properties of black hole systems are intimately connected to their radio properties \citep{koljonen15b}. Therefore, another observational link between the accretion flow and jet properties would not be surprising. However, an alternative explanation for the different radio/X-ray behavior for radio-loud and radio-quiet sources could be the source inclination, the effect of which to the radio and X-ray properties remains to be studied in the future.

\acknowledgments

We would like to thank the referees for their recommendations that have improved this paper. We also thank Piergiorgio Casella for insightful comments and discussion. This research has made use of data obtained from the High Energy Astrophysics Science Archive Research Center (HEASARC), provided by NASA's Goddard Space Flight Center.

%

\vspace{5mm}
\facilities{RXTE(PCA and HEXTE)}


\software{ISIS \citep{houck00}, Hyperfit \citep{robotham15}
          }

\bibliographystyle{aasjournal}

\bibliography{references}



\end{document}